\documentclass[reprint,superscriptaddress,onecolumn,showpacs,amsmath,amssymb]{revtex4-1}

\usepackage{graphicx}
\usepackage{dcolumn}
\usepackage{bm}
\usepackage{epsfig}
\usepackage{color}
\usepackage{longtable}

\usepackage{longtable}
\usepackage{color}

\begin{document}

\title{Relativistic effects on the hyperfine structures of $2p^4(^3P)3p \; ^2D^o, \; ^4D^o$ and $\; ^4P^o$ in $^{19}$F~I}

\author{Thomas Carette}
\affiliation{Chimie Quantique et Photophysique, Universit\'e Libre de Bruxelles, B-1050 Brussels, Belgium}
\affiliation{Department of Physics, Stockholm University, AlbaNova University Centre, SE-106 91 Stockholm, Sweden}

\author{Messaoud Nemouchi}
\affiliation{Laboratoire d'\'Electronique Quantique, Facult\'e de Physique, USTHB, BP32,
El-Alia, Algiers, Algeria}

\author{Jiguang Li}
\affiliation{Chimie Quantique et Photophysique, Universit\'e Libre de Bruxelles, B-1050 Brussels, Belgium}
\affiliation{Department of Physics, Lund University, S-221 00 Lund, Sweden}

\author{Michel Godefroid}
\email{mrgodef@ulb.ac.be}
\affiliation{Chimie Quantique et Photophysique, Universit\'e Libre de Bruxelles, B-1050 Brussels, Belgium}

\date{\today \\[0.3cm]}

\begin{abstract}
The hyperfine interaction constants of the $2p^4(^3P)3p$~$^2D^o_{3/2,5/2}$, $^4D^o_{1/2-7/2}$ and $^4P^o_{1/2-5/2}$ levels~in neutral fluorine are investigated theoretically. Large-scale calculations are carried out using the multiconfiguration Hartree-Fock (MCHF) and  Dirac-Hartree-Fock (MCDHF)  methods. In the framework of the MCHF approach, the relativistic effects are taken into account in the Breit-Pauli approximation 
using non relativistic orbitals. In the fully relativistic approach, the orbitals are optimized using the Dirac-Coulomb Hamiltonian with correlation models inspired by the non relativistic calculations. Higher-order excitations are captured through multireference configuration interaction calculations including the Breit interaction. In a third (intermediate) approach, the Dirac-Coulomb-Breit Hamiltonian matrix is diagonalized in a relativistic configuration space built with non relativistic MCHF radial functions converted into Dirac spinors using the Pauli approximation. The magnetic dipole hyperfine structure constants calculated with the three relativistic models are consistent and reveal unexpectedly large effects of relativity for $^2D^o_{5/2}$, $^4P^o_{3/2}$ and $^4P^o_{5/2}$. The agreement with the few available experimental values is satisfactory. The strong $J$-dependence of relativistic corrections on the hyperfine constants is investigated through the detailed analysis of the orbital, spin-dipole and contact relative contributions calculated with the non relativistic magnetic dipole operator.
 \end{abstract}

\maketitle

\section{Introduction} \label{sec:intro}

Atomic fluorine is a highly reactive free radical. Its natural state is molecular fluorine, a poisonous and corrosive material that makes experimental studies quite delicate and scarce~\cite{Musetal:99a,Shietal:2001a}. The atomic resonance transitions lie in the ultraviolet region but many of the transitions between excited states lie in the visible and near-infrared and can be driven using diode lasers, as explored by Tate and Aturaliye~\cite{TatAtu:97a} who reported for the first time high-resolution laser spectroscopy measurements of hyperfine structures. 
These authors used Doppler-free saturation absorption spectroscopy of excited states of atomic fluorine to measure and analyze the hyperfine structure intervals of the $2p^4 (^3P) 3s \; ^2P_J \rightarrow 2p^4 (^3P) 3p \; ^2D^o_{J'}$ fine structure multiplet components.  Using the observed hyperfine structure splittings~(hfs),  the magnetic hyperfine constants $A_J$ were determined for the levels involved in the transitions, with a higher accuracy than those determined earlier by Lid\'en~\cite{Lid:49a} and Hocker~\cite{Hoc:78a}. The comparison of the experimental hfs reveals large discrepancies~\cite{TatAtu:97a}. For instance, hfs values for the splittings of the $2p^4 (^3P) 3p \; ^2D^o_{3/2}$ and $^2D^o_{5/2}$ states have been found to be negative in~\cite{Hoc:78a} while positive in  \cite{Lid:49a} and \cite{TatAtu:97a}, with a large discrepancy (21 \%) between the two latter for the hfs values of $2p^4 (^3P) 3p \; ^2D^o_{3/2}$, well outside the error bars.
In contrast with the hyperfine study of the ground state levels $2p^5 \; ^2P^o_{1/2,3/2}$ for which observation~\cite{Radetal:61a,Har:65a,LagBea:82a} and theory~\cite{Besetal:63a,Schetal:69a,GlaHib:78b} have been compared, there is no theoretical prediction for the hyperfine structure of the excited levels considered in~\cite{TatAtu:97a}, except the pioneer work by Brown and Bartlett~\cite{BroBar:34a}.
More recently, in a feasibility study of in-beam polarization of fluorine, Levy {\it et al.}~\cite{Levetal:2007a} measured the hyperfine structures of $2p^4 (^3P) 3s \; ^4P_{5/2}$ and $2p^4 (^3P) 3p \; ^4D^o_{5/2,7/2}$ states via laser-induced fluorescence and modulated optical depopulation pumping.  As for $2p^4 (^3P) 3p \; ^2D^o_{3/2}$ and $^2D^o_{5/2}$, no theoretical values are available in the literature for these quartet levels, to the knowledge of the authors. \\

The present work was originally motivated by the following observation: on one hand, a serious  disagreement appeared when comparing our first theoretical estimation of the hyperfine constant of $2p^4(^3P)3p~^{2}D^o_{5/2}$ based on robust non relativistic calculations 
with the Doppler-free spectroscopy value reported by Tate and Aturaliye~\cite{TatAtu:97a}. On the other hand, ab initio calculations of hyperfine constants for $^{14}$N and $^{15}$N~\cite{Jonetal:2010a} were found to be in complete disagreement with the experimental values of Jennerich et al~\cite{Jenetal:06a}, also deduced from the analysis of the near-infrared Doppler-free saturated absorption spectra. This nitrogen theory-observation discrepancy problem was recently solved through a reinterpretation of the recorded weak spectral lines as crossover signals~\cite{Caretal:10b}, leading to a new set of experimental hyperfine constants in very good agreement with the ab initio predictions~~\cite{Jonetal:2010a}. Considering that the apparition of crossover signals in Doppler-free saturated absorption spectroscopy that has been used for both fluorine~\cite{TatAtu:97a} and nitrogen~\cite{Jenetal:06a}, is helpful in some cases but also  problematic in others, we investigate in the present work the relativistic corrections that could explain the non relativistic theory-observation discrepancy mentioned above for the $A_{5/2}$-value of $2p^4(^3P)3p~^{2}D^o_{5/2}$.

For light atomic systems, the relativistic effects are usually included with success in the Breit-Pauli approximation~\cite{FBJ:97a,FroTac:04a} for fine structure and transition probability calculations. In the case of fluorine, the relativistic corrections are expected to be relatively small. We expect therefore that relativity could be treated in a perturbation regime using either the Breit-Pauli approximation~\cite{ArmFen:74a,DyaFae:2007a} or the relativistic configuration interaction approach in the Pauli approximation~\cite{Joh:2007a,Gra:2007a}.  It is worthwhile to investigate if these methods lead to hyperfine structure constants consistent with each other, with the fully relativistic approach and with observation, when available. 
The evaluation of hyperfine interaction structures for atomic states provides a good opportunity to study the interplay between the correlation and relativistic effects. Different theoretical approaches can be used for estimating hyperfine structures, with their advantages and disadvantages, depending on the size and complexity of the targeted atomic systems. Fluorine has a special place in this diversity. As a nine-electron atom, it definitely lies outside the ``few''-electron systems domain for which the elaborate variational calculations in Hylleraas coordinates  can be successfully applied, usually giving rise to the most reliable expectation values~\cite{WuDra:2007a,Kin:2007a,PucPac:2009a}. Moreover, taken in its  $2p^4 3p$ excited configuration, neutral fluorine consitutes a difficult target for many-body approaches that are often restricted to single- or two-valence atoms or ions \cite{Paletal:2007a,SafJoh:2007a,Saf:2011a}. The coupled-cluster theory is promising \cite{SafSaf:2011a,
ManAng:2011a,Dutetal:2013a} but investigation of hyperfine structures in more complex  systems remains scarce \cite{Dasetal:2011a}. 
Although further developments might be expected \cite{Koz:2004a,DzuFla:2007a,Safetal:2009a,Veretal:2013a}, the traditional multi-configuration methods combined with configuration interaction in their non relativistic \cite{Jonetal:2010a,CarGod:2011a,CarGod:2011b} and relativistic \cite{Maletal:2001a,NorBec:2001a,Bieetal:2009a,JonBie:2010a,Jonetal:2010b} versions keep a respectable place in the ranking of ab initio methods for hyperfine structures calculations.

 Section~\ref{sec:AS} describes the atomic state functions in the non relativistic multiconfiguration Hartree-Fock, relativistic Breit-Pauli, relativistic multiconfiguration Dirac-Hartree-Fock and Pauli approximations. The theoretical background needed for understanding the hyperfine interaction is given in Section~\ref{sec:HI}. The computational strategy is developed in Section~\ref{sec:CM}. The theoretical hyperfine constants calculated using the different models are compared to each other and with observation  in Section~\ref{sec:RD}.

\newpage

\section{The Atomic State Function} \label{sec:AS}
\subsection{In the non relativistic approach}

In the non relativistic multiconfiguration Hartree-Fock (MCHF) approximation~~\cite{Fro:77a}, the atomic state function~(ASF) is described as a linear combination of $N_c$ configuration state functions (CSFs)
\begin{equation}
\label{MCHF_wfn}
\Psi (\alpha  L S M_L M_S  \pi ) = \sum_i^{N_c} c_i \; \Phi( \alpha_i L S M_L M_S \pi) 
\end{equation}
built on one-electron spin-orbitals
\begin{equation}
\label{nr_spin_orbitals}
\phi_{nlm_lm_s}({\bf r}, \sigma ) = \frac{P_{nl}(r)}{r} Y _{l m_l} (\theta, \varphi) \chi_{m_s} (\sigma) \; .
\end{equation}
 The MCHF equations are the system of coupled, non-linear differential equations that arise when we require the energy to be stationary with respect to variations in the radial functions $\{ P_{nl} (r) \}$. At the same time the energy must be stationary with respect to variations in the mixing coefficients $\{ c_i \}$, leading to a system of secular 
equations~\cite{FBJ:97a}.
Once a set of one-electron orbitals optimised, a larger system of secular equations can be solved for diagonalizing the non relativistic Hamiltonian in an enlarged CSF basis to get a better description of the desired eigenvector. In the present paper, we will refer to these calculations  as configuration interaction (CI).

\subsection{In the relativistic approach} \label{ssec:RC}

\subsubsection{Breit-Pauli approximation}

Relativistic corrections to the MCHF or CI wave functions can be included efficiently in the Breit-Pauli (BP) approximation~\cite{FBJ:97a} that consists in writing the ASF as the following  expansion
\begin{equation}
\label{BP_wfn}
\Psi (\alpha  J M_J \pi ) = \sum_i^{N_c} c_i \; \Phi( \alpha_i L_i S_i J M_J \pi ) \; .
\end{equation} 
This wave function is the eigenvector of the Breit-Pauli Hamiltonian matrix corresponding to the desired root. Note that, oppositely to (\ref{MCHF_wfn}), the ASF (\ref{BP_wfn}) allows $LS$-mixing due to the fine-structure BP Hamiltonian terms that do not commute with ${\bf L}$ and ${\bf S}$~\cite{Hibetal:91a,FBJ:97a}.

\subsubsection{Multiconfiguration Dirac-Hartree-Fock approach}

Starting from the Dirac-Coulomb Hamiltonian~\cite{Gra:2007a}
\begin{eqnarray}
\label{DC_Ham}
H_{DC} = \sum_{i}^N \left( c\, {\bm{ \alpha }}_i \cdot {\bm{ p }}_i + (\beta_i -1)c^2 + V^{nuc}_i \right) + \sum_{i>j} 1/r_{ij},
\end{eqnarray}
where $V^{nuc}$ is the monopole part of the electron-nucleus Coulomb interaction, the atomic state function (ASF) describing a specific fine structure level is described by a linear combinations of relativistic configuration state functions $\Phi(\gamma_{i}  J M_J \pi)$ 
\begin{eqnarray}
\label{jj_ASF} 
\Psi(\gamma J M_J \pi)  = \sum_{i=1}^{N_c} c_{i} \; \Phi(\gamma_{i}  J M_J \pi).
\end{eqnarray}
that are built on relativistic configurations $\gamma_i$ involving the $jj$-coupling of subshell  Dirac spinors~\cite{Gra:2007a} 
\begin{equation}
\label{Dirac_spinor}
\phi_{n \kappa m} ({\bf r}, \sigma )  = \frac{1}{r} \left( 
\begin{array}{c}
P_{n \kappa} (r) \; \chi_{\kappa m} (\theta , \varphi) \\
i Q_{n \kappa} (r) \; \chi_{- \kappa m} (\theta , \varphi) 
\end{array}
\right) \; ,
\end{equation}
where  $\kappa$ is defined as
\begin{eqnarray}
\kappa = \left\{\begin{array}{cl} -l-1 &\qquad \textrm{when} \qquad j=l+1/2\\
l &\qquad \textrm{when} \qquad j=l-1/2 \; . \end{array}\right.
\end{eqnarray}
Applying the variational principle, the radial functions $\{P_{n \kappa}(r), Q_{n \kappa}(r) \}$  and the mixing coefficients $c_i$ appearing in~(\ref{jj_ASF})  are optimized by solving iteratively the self-consistent field (SCF) problem and the secular equations. 
Calculations can be performed for a single level, but also for a portion of a spectrum in an extended optimal level (EOL) scheme where optimization is applied on a weighted sum of energies. In the Extended Optimal Level (EOL) optimization scheme \cite{Graetal:80a,Gra:2007a} that we adopt for the present study (using the ``standard'' option of the {\sc GRASP2K}~\cite{Jonetal:2007a} computer code), the functional has the form
\begin{equation}
\label{EOL_fct}
{\cal F} = \sum_{r=1}^{N_c} \sum_{s=1}^{N_c} \; d_{rs} H_{rs} + \mathcal{L} \; ,
\end{equation}
where $\mathcal{L}$ contains the Lagrange multipliers contributions and
\begin{equation}
\label{EOL_d_rs}
d_{rs} = \left(  \sum_{i=1}^{n_L} (2J_i+1) c_{ri} c_{si} \right ) / \left(  \sum_{i=1}^{n_L} (2J_i + 1) \right) \; .
\end{equation}
 $n_L$ specifies the number of the targeted eigenvalues, each of them weighted by the $(2J_i+1)$ degeneracy factor.

An extension of the MCDHF approach, allowing  the mixing coefficients to be varied but keeping the one-electron orbitals frozen, is referred in the present work as the relativistic configuration interaction (RCI) method. In the latter, 
the  transverse photon interaction~\cite{Dyaetal:89a}
\begin{eqnarray}
\label{eq:Transv}
H_{Transv} = 
- \sum_{i < j}^{N} \left[ 
\boldsymbol{\alpha}_i\cdot\boldsymbol{\alpha}_j
\frac{ \cos\left(\omega_{ij}r_{ij}/c \right) }{r_{ij}}
+
(\boldsymbol{\alpha} \cdot \boldsymbol{\nabla})_i(\boldsymbol{\alpha} \cdot \boldsymbol{\nabla} )_j
\frac{\cos\left(\omega_{ij}r_{ij}/c \right)-1}{\omega_{ij}^2 r_{ij}/c^2} \right] \; ,
\end{eqnarray}
may be  included in the Hamiltonian matrix. 
However $\omega_{ij}$ appearing in this equation is the energy of the exchanged photon between the two electrons $(i,j)$,  and  is   not well defined for correlation orbitals. Therefore, it is only possible to estimate the low-frequency $(\omega_{ij} \rightarrow 0)$ limit of (\ref{eq:Transv}) by multiplying the computed photon frequency by a small number  to get the Breit interaction~\cite{Dyaetal:89a,Paretal:96a}
\begin{eqnarray}
\label{eq:Breit}
         H_{Breit} =   -\sum_{i < j}^{N}\frac{1}{2 r_{ij}}\Biggl[ \bm{\alpha}_{i} \cdot \bm{\alpha}_{j} + \frac{ (\bm{\alpha}_{i} \cdot {\bm{ r_{ij} }})
                (\bm{\alpha}_{j} \cdot {\bm{ r_{ij} }}) } {r_{ij}^2} \Biggr] \; .
\end{eqnarray}

\subsubsection{Pauli approximation}

Another interesting way to estimate relativistic effects  is to diagonalise the Dirac-Coulomb-Breit Hamiltonian 
$( H_{DC} + H_{Breit})$ matrix,   in a relativistic CSF basis built on  Dirac spinors whose large and small radial components are calculated from non relativistic  
MCHF radial functions, using the Pauli approximation~\cite{Gra:94a,Joh:2007a,Gra:2007a}
\begin{eqnarray}
\label{Pauli_approx}
P_{n\kappa}(r) &=& P_{nl}^{MCHF}(r) \; ,\nonumber \\
Q_{n\kappa}(r) & \approx & \frac{\alpha}{2} \left( \frac{d}{dr} + \frac{\kappa}{r} \right) P_{n \kappa}(r) \; ,
\end{eqnarray}
where $\alpha$ is the fine structure constant.  This method based on the use of the relativistic configuration interaction approach in the Pauli approximation is labelled RCI-P in the present work.

\section{Hyperfine Interaction}
\label{sec:HI}

The hyperfine contribution to the Hamiltonian is  represented by a multipole expansion 
\begin{equation}
\label{H_hfs}
  H_{\rm{hfs}}=\sum_{k\geq1} {\bf T}^{(k)} \cdot {\bf M}^{(k)}
\end{equation}
where ${\bf T}^{(k)}$ and ${\bf M}^{(k)}$ are spherical tensor operators of rank $k$ in the electronic and nuclear space, respectively~\cite{Sch:55a,Arm:71a}. The $k=1$ and $k=2$ terms 
represent, respectively, the magnetic dipole interaction and the electric quadrupole interaction. The $^{19}$F nucleus, the only stable fluorine isotope, has a nuclear spin $I=1/2$ and a magnetic moment $\mu_I = 2.628868~\mu_N$~\cite{Rag:89a,Sto:05a,Sto:2011a} but no quadrupole moment ($Q=0$) . 
The hyperfine shifts of the fine-structure levels may be expressed to first order in terms of the magnetic dipole $A_J$  hyperfine interaction constant~\cite{Jonetal:93a} that is  proportional to the reduced matrix element of the electronic tensor operator of rank one
\begin{equation}
\label{A_J_RME}
A_J=\frac{\mu_I}{I}\frac{1}{\sqrt{J(J+1)(2J+1)}}\langle\gamma J \Vert {\bf T}^{(1)} \Vert \gamma J\rangle \; .
\end{equation}
In non relativistic calculations, the electronic  matrix elements are obtained by integrating
the irreducible spherical tensors~\cite{LinRos:74a,Hib:75b} 
\begin{eqnarray}
 \mbox{\bf T} ^{(1)} = \frac{\alpha^2}{2} \sum_{i=1}^{N}
   \big\{ 2 \mbox{\bf l}^{(1)} (i) r_i^{-3}
  -g_s \sqrt{10} [ \mbox{\bf C}^{(2)}(i) \times \mbox{\bf s}^{(1)}(i) ] ^{(1)} r_i^{-3} 
  + g_s \frac{8}{3} \pi \delta( \mbox{\bf r}_i ) \mbox{\bf s}^{(1)}(i) \big\}
\label{T_1}
\end{eqnarray}
using the ASF of the form~(\ref{MCHF_wfn}) adapted to the ${\bf J} = {\bf L} + {\bf S}$ symmetry
\begin{equation}
\label{MCHF_J}
\Psi ( \alpha L S J M_J \pi ) = \sum_i^{N_c} a_i \; \Phi( \alpha_i L S J M_J \pi) \; ,
\end{equation}
i.e. an expansion similar to (\ref{BP_wfn}), but restricted to the same $LS$-values.
For light atoms in which the $LS$ coupling remains valid to a good approximation, relativistic corrections can be introduced in the Breit-Pauli (BP) approximation. The resulting wave functions (\ref{BP_wfn}) used to evaluate the matrix elements of the electronic tensor operator~(\ref{T_1}) allow  $LS$-mixing for a specific $J$-value.
In both cases, the hyperfine constant defined by (\ref{A_J_RME}) is composed of the orbital, spin-dipole and contact contributions 
 \begin{equation}
\label{A_J_split}
A_J = A_{J}^{orb} + A^{sd}_{J} + A^{c}_{J} \; ,
\end{equation}
that are evaluated using the eigenvectors (\ref{MCHF_wfn}) or (\ref{BP_wfn}). 
In cases where $LS$ coupling is strictly valid, i.e. omitting the ($L' \neq L$) and ($S' \neq S$) off-diagonal relativistic matrix elements,
 the three contributions to the hyperfine constant appearing in (\ref{A_J_split}) take the form
\begin{eqnarray}
\label{A_contributions}
\label{sec3:A_orb}
  A_{J}^{orb} & = & G_{\mu}\frac{\mu_{I}}{I}\;a_{l} \; {\cal F}^{orb}(L,S,J) \; , \\
 \label{sec3:A_sd}
A^{sd}_{J} &= &\frac{1}{2}\,G_{\mu}\,g_{s}\,\frac{\mu_{I}}{I}\,a_{sd} \; {\cal F}^{sd}(L,S,J) \; , \\
\label{sec3:A_c}
A_{J}^{c} &=&\frac{1}{6}\,G_{\mu}\,g_{s}\,\frac{\mu_{I}}{I}a_{c} \; {\cal F}^{c}(L,S,J) \; ,
\end{eqnarray}
where the  $J$-independent orbital~($a_{l}$), spin-dipole~($a_{sd}$) and contact~($a_{c}$)  electronic hyperfine parameters are defined as~\citep{LinRos:74a,Hib:75b}
\begin{align}
\label{eq3:small_a_l}
a_{l}&\equiv\langle \alpha LS(M_{L}=L)(M_{S}=S)|\sum_{i=1}^{N}
l^{(1)}_{0}(i)r^{-3}_{i}|\alpha LS (M_{L}=L)(M_{S}=S) \rangle \; ,  \\
\label{eq3:small_a_sd}
a_{sd}&\equiv\langle \alpha LS(M_{L}=L)(M_{S}=S)|\sum_{i=1}^{N}
2C^{(2)}_{0}(i)s^{(1)}_{0}(i)r^{-3}_{i}|\alpha LS(M_{L}=L)(M_{S}=S)\rangle  \; , \\
\label{eq3:small_a_c}
a_{c}&\equiv\langle \alpha LS(M_{L}=L)(M_{S}=S)|\sum_{i=1}^{N}
2s^{(1)}_{0}(i)r^{-2}_{i}\delta(r_{i})|\alpha LS(M_{L}=L)(M_{S}=S)\rangle  \; .
\end{align}
The dimensionless factors ${\cal F}^i (L,S,J)$ can be evaluated from the following expectation values
\begin{eqnarray}
\label{sec3:F_orb}
  {\cal F}^{orb}(L,S,J) & = & \frac{\langle\,\pmb{L}\cdot\pmb{J}\,\rangle}{LJ(J+1)} \; , \\
 \label{sec3:F_sd}
{\cal F}^{sd}(L,S,J)  & = &\frac{3\,\langle\,\pmb{L}\cdot\pmb{S} \rangle\, \langle\,\pmb{L}\cdot\pmb{J}\,\rangle\,-\,L(L+1)\,\langle\,\pmb{S}\cdot\pmb{J}\,\rangle}{SL(2L-1)J(J+1)}     \; , \\
\label{sec3:F_c}
{\cal F}^{c}(L,S,J)  &=& \frac{\langle\,\pmb{S}\cdot\pmb{J}\,\rangle}{SJ(J+1)}  \; .
\end{eqnarray}
Expressing the electronic parameters $a_l$, $a_{sd}$ and $a_c$ in atomic units (units of $a_0^{-3}$) and $\mu_I$ in nuclear magnetons ($\mu_N$), the magnetic dipole hyperfine structure constants $A_J$ are calculated in units of frequency (MHz) by using $G_\mu = 95.41067$. 

In fully relativistic calculations,  the structure of the magnetic dipole electronic tensor is much simpler than the non relativistic form~(\ref{T_1})~\cite{LinRos:74a,Jonetal:96c} 
\begin{equation}
\label{T1_rel}
 \mbox{\bf T} ^{(1)} = -i \alpha \sum_{j=1}^{N} 
 \left( {\bm{ \alpha }}_j \cdot
  \mbox{\bf l}_j \;  \mbox{\bf C}^{(1)}(j)  \right)  \frac{1}{ r_j^2} \; .
\end{equation}
The hyperfine constant $A_J$ is estimated from the expectation value of this operator, using (\ref{A_J_RME}) and the 
atomic state function~(\ref{jj_ASF}).

\section{Computational Models} \label{sec:CM}

\subsection{Non-relativistic calculations}

We  perform two types of non relativistic calculations. 
The first one is based on the multiconfiguration Hartree-Fock approach~\cite{FBJ:97a,Froetal:2007a} with configuration
expansions generated by single (S) and double (D)  excitations from the single reference. 
For a given calculation, the orbital active space (AS) is characterized by $[n_{max}]$ when no angular limitation applies. 
 The active set is specified by $[n_{max} l_{max}]$ when angular orbital limitation  is introduced. We have performed systematic SD-MCHF calculations, considering angular momentum values up to $l=5$ ($h$-electrons), and concluded that truncating the AS at $l_{max}=3$ is safe for getting hyperfine constants within $0.2$~\%. 
These calculations are denoted SD-MCHF in Table~\ref{tab:A_J_MCHF_BP}. 

With these MCHF orbital sets, we investigate the use of SD-multirefence expansions by performing configuration interaction calculations (MR-CI) based on expansions generated by allowing SD excitations from the three configurations $\{2s^2 2p^43p,~ 2s^2 2p^23p3d^2,~ 2s2p^43p3d \}$. To keep the size of the interaction matrices manageable, the three CSFs of the MR space are not treated identically in terms of SD excitations, considering a smaller orbital active set $[6f]$ for the two $\{ 2s^2 2p^23p3d^2,~ 2s2p^43p3d \}$ components than the one adopted $[10f]$ for the major ($2s^2 2p^43p$) component.
We observe that the use of a multireference space is worthwhile, bringing a $ 3.6$\% variation in the hyperfine constants of the $\; ^4P^o_{1/2}$ level.
For the $\; ^4D^o$ symmetry, we  test a ``reduction strategy'' that consists in limiting in the final expansions the excited CSFs that interact with at least one of the three MR components. These calculations are performed using  {\sc lsreduce} that is part of the utilities provided in the MCHF atomic-structure package~\cite{Froetal:2007a} and are labelled MR-CI-red in the present work. While the  number of CSFs is sensitively decreased by this reduction strategy (from 394~190 to 206~340), the hyperfine constants are not affected, as illustrated by Table~\ref{tab:A_J_MCHF_BP}.

The use of the multireference is also tested in the orbital optimization by performing MR-MCHF calculations to capture higher-order correlation effects. For the latter, we use the above reduction strategy, adopting the reversed orbital order consisting in coupling sequentially the subshells by decreasing $n$ and $l$. This technique indeed reduces substancially the size of the MCHF expansions while keeping the dominant correlation contributions~\cite{Jonetal:2010a}. 
 For specifying the AS, it is sometimes more convenient to use another notation involving curly brackets instead of brackets, where the number of orbitals for each angular symmetry is specified, i.e.
$\{ 10s9p8d7f \} = [10f]$. 
The $\{ 10s9p8d4f \}$ AS  used for the MR-MCHF calculations   means that for its three multireference components, the orbital angular momentum is limited to $l_{max}=3$ for $n \leq 7$ and $l_{max}=2$ for $n \geq 8$. As shown by  Table~\ref{tab:A_J_MCHF_BP}, the inclusion of the MR in the MCHF model reproduces the MR-CI  results within less than 1~\%.

\subsection{Relativistic calculations}

In the Breit-Pauli (BP) approximation, the CSF expansions of the atomic state function (\ref{BP_wfn}) are constructed in the same way than in the SD-MCHF calculations,  but including all possible symmetries $L_i S_i$ for a given $J$-value. The radial functions spanning the CSFs are taken from the $[nf]$ SD-MCHF calculations. All the Breit-Pauli operators are taken into account. 

Relativistic configuration interaction calculations are also performed in the Pauli approximation (RCI-P) by generating $jj$-coupled relativistic configuration state function expansions (\ref{jj_ASF}) from SD excitations of the monoreference configuration using the $[10f]$ active set. The radial functions are the non relativistic MCHF radial functions converted to approximate Dirac spinors according to (\ref{Pauli_approx}). 

Replacing the monoreference by a MR model  in the non relativistic framework brings variations of a few percents in the hyperfine constants, as shown in the previous section. It is therefore worthwhile  investigating the multireference approach in the Breit-Pauli approximation.
For these calculations (denoted MR-BP), we build the CSF expansions by including  SD excitations from the three configurations $\{2s^2 2p^43p,~ 2s^2 2p^23p3d^2,~ 2s2p^43p3d \}$ multireference, using respectively the $[10f]$, $[5d]$ and $[5d]$ active space. For a given $J$-value, all symmetries resulting from the main reference ($2s^2 2p^43p$) are  included, while for the other two references $\{ 2s^2 2p^23p3d^2,~ 2s2p^43p3d \}$ only the $^2(S,P,D)$ and $^4(S,P,D)$ symmetries are considered. The size of the spaces are reduced with {\sc lsreduce}. \\

The three sets of BP,  RCI-P and MR-BP results are presented in the second half of Table~\ref{tab:A_J_MCHF_BP}.
 \\

In Table~\ref{tab:A_J_MCDHF}, we report fully relativistic results.
In the non relativistic  approximation, the desired states $2p^43p \; ^4D^o$, $^4P^o$, $^2D^o$ are the lowest  of their symmetry. This is not true anymore in the relativistic framework for the $J=1/2,3/2$ levels for which the interaction with the ground configuration $2p^5$ should be taken into account. The simplest model is therefore a two-configuration model $\{ 2p^43p + 2p^5 \}$ for these $J$-subspaces. MCDHF calculations are performed by using the 
 active space  approach inspired from the non relativistic SD-MCHF correlation models. Denoting  the $n^{\mbox{th}}$ root of the $J$-block by $E(n_J)$ and referring to (\ref{EOL_fct}) and (\ref{EOL_d_rs}), the EOL strategy is applied to optimize separately three orbitals sets, using the
  following energy functionals
\begin{itemize}
\item $[6 E(1_\frac{5}{2}) + 4 E(2_\frac{3}{2}) + 2 E(2_\frac{1}{2}) ] / 12 $,
\item $ [8 E(1_\frac{7}{2}) + 6 E(2_\frac{5}{2}) + 4 E(3_\frac{3}{2}) + 2 E(3_\frac{1}{2}) ] / 20 $,
\item $ [6 E(3_\frac{5}{2}) + 4 E(4_\frac{3}{2}) ] / 10 $,
\end{itemize}
describing respectively the fine structure levels $J$ of the three terms  $2p^4(^3P)3p~^4P^o, ~^4D^o, ~^2D^o$.

 The number of CSFs in SD-MCDHF expansions increases drastically with the extension of the AS compared to SD-MCHF ones. To keep the size of the multiconfiguration expansions manageable the reduction strategy, that has been proven to be efficient in the non relativistic MR-CI calculations for the $^4D$ symmetry, is applied by using the {\sc jjreduce} code~\cite{Jonetal:2013a}. Moreover, the orbital active sets are restricted to $l_{max}=2$ from  $n=8$, as indicated by the curly bracket notation used in  Table~\ref{tab:A_J_MCDHF}. Another difference with the SD-MCHF strategy is that the calculations  are carried out layer by layer, i.e. optimizing only the correlation orbitals of the added layer together with the mixing coefficients.  
 The Breit interaction~(\ref{eq:Breit}) is taken into account in the subsequent RCI computations. The configuration space is built by allowing SD excitations from the same multireference $\{2s^2 2p^43p,~ 2s^2 2p^23p3d^2,~ 2s2p^43p3d \}$ as the one used in the non relativistic calculations. To keep the configuration interaction problem tractable, we adopt two different active spaces: $\{10s9p8d3f\}$ for the major component $2s^2 2p^43p$ and $[5d]$ for the two others. Moreover, for  $2s^2 2p^23p3d^2 $, the excitations are restricted to the ones in which the $1s$ shell remains closed.

\section{Results and discussion}
\label{sec:RD}

The convergence of the hyperfine structure constants with the progressive extension of the orbital active sets within a given correlation model is satisfactory, as illustrated by Tables~\ref{tab:A_J_MCHF_BP} and \ref{tab:A_J_MCDHF} for the non relativistic SD-MCHF and relativistic SD-MCDHF results, respectively.
The  excellent agreement between the BP and the RCI-P results is rather comforting. Both sets  arise from the same radial one-electron orbitals optimized through the non relativistic MCHF approach but relativity is included not only through different approaches but also using independent computational tools ({\sc ATSP2K}~\cite{Froetal:2007a} and {\sc GRASP2K}~\cite{Jonetal:2007a} codes).
The Breit-Pauli Hamiltonian is indeed a low-order approximation of the Dirac-Coulomb Hamiltonian 
and the expectation values of its operators are evaluated using non relativistic $LSJ$-basis functions while the RCI-P  method diagonalizes the Dirac-Coulomb-Breit Hamiltonian in a $jjJ$-CSF basis built on approximated Dirac spinors. Moreover the evaluation of the expectation values of the magnetic dipole electronic  tensors $(\ref{T_1})$ and $(\ref{T1_rel})$ is done within radically different frameworks.
The effect of enlarging the reference set that can be estimated by comparing the MR-CI and SD-MCHF values for a given active set $([10f]$ is coherent with the MR-BP and BP differences found in the Breit-Pauli approximation. This means that enlarging the multi-reference space mostly captures electron correlation.
A detailed cross-comparison of the most elaborate calculations reported in Tables~\ref{tab:A_J_MCHF_BP} and \ref{tab:A_J_MCDHF} shows that enlarging the reference space improves the agreement between the Breit-Pauli and fully relativistic values.

 We present in Table~\ref{tab:A_J_comparison} the magnetic dipole hyperfine constants corresponding to the largest AS for each theoretical model and compare them with experimental values when available.  
As already observed above, the two sets of non relativistic MR-MCHF and MR-CI values are consistent with each other, but the comparison with the SD-MCHF values indicates the significant effect of higher-order excitations.
In the mono-reference model, the comparison between the SD-MCHF and BP values reveals the importance of the relativistic corrections for some levels. This is a priori unexpected for a light system such as neutral fluorine. Amongst the nine levels considered, the hyperfine constants of  $^4P^o_{3/2}$ and $^4P^o_{5/2}$ are the most affected by relativity, the difference between the BP and SD-MCHF results reaching as much as 30~\%. This effect is less important but still quite large for $^2D^o_{5/2} (17\%) $, $^4D^o_{3/2} (10\%) $ and $^4P^o_{1/2} (7\%) $. The same observation can be made from the relativistic configuration interaction calculations in the Pauli approximation (RCI-P).

The fully relativistic results  (MCDHF) confirm the large relativity effects found in the Breit-Pauli approximation. The comparison between the MCDHF and BP/RCI-P values, all based on mono-reference correlation models, is by itself interesting,  illustrating the rather good coherence (within 2~\%) of the hyperfine constant values.
The agreement between the Breit-Pauli  and fully relativistic values is maintained  when enlarging the reference space. The agreement between MR-BP and MR-RCI is indeed better than 1.8\%. 
Going from MCDHF to MR-RCI, one takes into account, not only the higher-order excitations beyond the monoreference model (including the interplay between electron correlation and relativity), but also the Breit interaction (\ref{eq:Breit}). The corresponding variation systematically improves the theory-observation agreement in the four $A_J$-values for which experimental data are available~\cite{Levetal:2007a,TatAtu:97a}.  The remaining discrepancies between experiment and theory arise most likely from higher-order electron correlation. 
Unfortunately, experimental values are limited to four levels amongst the nine considered. Taking these values as the truth, the uncertainty of the (MR-BP/MR-RCI) average values is estimated to be better than 3\%. With this respect, the 5\% difference between theory and observation for $\; ^2D^o_{3/2}$ is somewhat surprising, as suggested by the following detailed analysis.
\\

To get some insight in the origin of the strong level-dependence of relativistic effects, we  report in Table~\ref{tab:A_J_ventilation}  the  SD-MCHF and BP values of the three different  hyperfine contributions $A_{i}$ ($i=orb, sd, c$).
 The ratios of the SD-MCHF values, for a given contribution $i$, are strictly conditioned by the 
 factors ${\cal F}^{orb} (L,S,J)$  ,  ${\cal F}^{sd} (L,S,J)$ 
 and  ${\cal F}^{c} (L,S,J)$  defined in equations~(\ref{sec3:F_orb}), (\ref{sec3:F_c}) and~(\ref{sec3:F_sd}). These are 
 explicitly reported in Table~\ref{tab:F_factors}.
 For instance, the first line numbers $(4360:1744:1370:1246)$ appearing in Table~\ref{tab:A_J_ventilation} and corresponding to the $A_{orb}$ contributions of $^4D^o_J$ are in the ratios $(35:14:11:10)$ that can be found in the  ${\cal F}^{orb}$ column of Table~\ref{tab:F_factors}. Similarly, the sixth line numbers $(-497:676:-179)$ reporting the $A_{sd}$ values for $^4P^o_J$ in Table~\ref{tab:A_J_ventilation}  follow the ${\cal F}^{sd}$ ratios $(50:-68:18)$ of Table~\ref{tab:F_factors}. As can be realized from the BP values reported in Table~\ref{tab:A_J_ventilation}, these ratios are strongly affected by the relativistic corrections in the Breit-Pauli approximation due to the $LS$-mixing in (\ref{BP_wfn}).
For example, the ratios $(35:14:11:{\it 10})$ and $(50:{\it -68}:18)$ calculated from the corresponding $A_{orb}$ and $A_{sd}$ MCHF values, respectively,  become $(32.3:12.5:10.3:{\it 10})$ and $(70.9:{\it -68}:5.4)$. 
 Table~\ref{tab:F_factors}  includes the $LS$ composition of the BP wave functions. The strongest $LS$-mixing appear for the $^2D^o_{5/2}$ and $^4D^o_{5/2}$ terms, but the purity of all $^4P^o_J$ levels remains high.  One can then conclude that there is no trivial correlation between the $LS$-mixing magnitude and the relativistic effect on the hyperfine constant value.  
 
 In the Table~\ref{tab:A_J_ventilation}, we  report the relative differences between the SD-MCHF and BP values
\begin{equation} \label{A_i_over_A_tot}
\frac { \Delta A_{i} }{A_{tot}} = \frac{ A_{i} \mbox{(SD-MCHF)} -A_{i} \mbox{(BP)} }{ A_{tot}\mbox{(SD-MCHF)} }
\end{equation}
 for the  three different  hyperfine contributions,  using A$_{tot}$(SD-MCHF) as the reference value.
The analysis of these relative contributions  sheds some light on the $J$-level-dependence of relativistic effects for a given $LS$ term.
 The 31\% found for the relativistic effect on $A_{5/2}$ of $2p^4(^3P)3p$~$^4P^o_{5/2}$ is due to the cumulative effects of $+14.5$\% and $+16.7$\% relativistic contributions to the orbital and spin-dipole contributions, while the very small impact of relativity ($-2.4$\%) found on 
$A_{1/2}$ of $2p^4(^3P)3p$~$^4D^o_{1/2}$  is explained by the strong cancellation of the (still large in absolute value) $-15.4$\% and $+12.9$\% relativistic contributions to the orbital and spin-dipole contributions.
 
\section{Conclusion}

Relativistic effects on the hyperfine structures of heavy elements are well known. Woodgate showed that a calculation of the breakdown of $L-S$ coupling and of second-order corrections, off-diagonal in $J$, is necessary for an interpretation of the spectrum of samarium~\cite{Woo:66a}. It has  been shown independently by Sandars and Beck \cite{SanBec:65a} that hyperfine structure calculations can often be made more conveniently by using effective operators between non relativistic $LS$ basis states. This approach has been used for instance by Childs \cite{Chi:71a} for studying relativistic effects in the hyperfine structure of the tin isotopes. A critical analysis of the methods used to interpret the hyperfine structure in complex free atoms and ions can be found in \cite{Demetal:2010a}.
The investigation of relativity on hyperfine parameters in light systems is less common. In the present work, robust correlation models are  built in the non relativistic approach, to investigate hyperfine structure parameters in fluorine. The reliability of these models is assessed by comparing single- and double-, mono- and multi-reference MCHF and CI calculations that all agree with each other within, at most, 3.5\%. For some levels -~$2p^4(^3P)3p$~$^2D^o_{5/2}$ is a nice example -,  all non relativistic correlation models perfectly agree with each other but  differ quite substantially ($\simeq 17$\%) from observation.
It is well known that relativistic effects on the electronic atomic structures are growing with the nuclear charge~\cite{Gra:2007a,Joh:2007a} but are expected to be smaller than electron correlation for neutral and light atomic systems. 
In neutral fluorine,  yet a very light element $(Z=9)$, we show that relativistic corrections to the non relativistic hyperfine parameters can be large for some low-lying levels, reaching around 30\% for the $A$-values of 
$2p^4(^3P)3p$~$^4P^o_{3/2}$ and $^4P^o_{5/2}$.
While non-relativistic approaches are often successful in computing hyperfine constants with good accuracy, even in heavier systems~\cite{CarGod:2011b}, we see here that it is necessary to systematically estimate relativistic corrections. In this context, BPCI and RCI-P methods stand as valuable tools since they are computationally cheap compared to fully relativistic calculations.

Core-orbital contraction and charge density rearrangement due to relativity can be very important \cite{LinRos:74a} and are  a priori poorly described in the MCHF-BP approximation~\cite{JonBie:2010a}. For fluorine however, the hyperfine structure parameters estimated with the MCHF-BP method are nicely coherent with the results obtained from the fully relativistic MCDHF method, suggesting that the orbital contraction effects are minor in comparison to the $LS$ term relativistic mixing. When both methods produce similar results, the first approach (MCHF-BP) offers some advantages  compared with the second one (MCDHF). 
The analysis of the relative orbital, spin-dipole and contact contributions, that is difficult in the MCDHF framework~\cite{Pyp:88a,ZhaPyp:88a} and that becomes impossible when using the simple form of the magnetic dipole operator~(\ref{T1_rel}), sheds indeed some light in the origin of the large $J$-dependency of relativistic effects, as we explicitly illustrate in the present work.

Refining our preliminary non relativistic results by introducing relativity through the Breit-Pauli Hamiltonian, we find large unexpected variations in the hyperfine structure constants that evidently bring the theoretical estimations closer to the experimental values of Tate and Aturaliye~\cite{TatAtu:97a}. 
While it has been clearly demonstrated that the theory-observation disagreement was due to a wrong interpretation of the Doppler-free saturated absorption spectroscopy signals in nitrogen~\cite{TatAtu:97a}, a  good agreement is found with the fluorine experimental values obtained with the same technique if the relativistic corrections are included. This observation excludes any misinterpretation of the crossover signals in fluorine. 
We identify in the present work the origin of the relativistic effects on the hyperfine constants for specific levels and expect them to be even larger for levels that are not yet considered experimentally. We are strongly encouraging experimental studies of the hyperfine structures in fluorine to confirm our theoretical estimation on the crucial role of relativity, in particular for the $^4P^o_{3/2,5/2}$ levels.

\begin{table}
\caption{Hyperfine structure constants $A_J$ (in MHz) of $2p^4(^3P)3p \; ^4D^o_{J}$, $\; ^4P^o_{J}$ and $\; ^2D^o_{J}$. Upper part: non relativistic values obtained with the multiconfiguration Hartree-Fock method using single- and double-monoreference expansions (SD-MCHF), and multireference configuration interaction (MR-CI) calculations. Lower part: relativistic values calculated in the Breit-Pauli (BP and MR-BP) and the Pauli (RCI-P) approximations.  }
\begin{center}
\begin{tabular}{lccccccccccccc} \\
 \\
\hline \hline \\[-0.2cm]
  &          && \multicolumn{4}{c}{$^4D^o$} && \multicolumn{3}{c}{$^4P^o$} && \multicolumn{2}{c}{$^2D^o$ }  \\
Method & AS      &&  $A_{1/2}$ & $A_{3/2}$ & $A_{5/2}$ & $A_{7/2}$ && $A_{1/2}$ & $A_{3/2}$ & $A_{5/2}$ && $A_{3/2}$ & $A_{5/2}$ \\ [0.1cm]
\hline  \\[-0.2cm]
HF &       &&  2466      & ~987       & 1109      & 1538      &&  $-$1991     &  1260     &  731      &&   1862     &  2037     \\ 
 &          &&            &           &           &           &&            &           &           &&            &           \\
SD-MCHF & [3]         &&  1535    & 1152      & 1433      & 1919      &&  $-$~454      &  1935     & 1274      &&   1509     &  2387   \\
& [4]         &&  2275      & ~893       & 1033      & 1474      &&  $-$1961     &  1226     &  674      &&   1768     &  1954     \\
& [5]         &&  2087      & ~931       & 1113      & 1579      &&  $-$1820     &  1330     &  744      &&   1705     &  2058     \\
& [$6f$]        &&  2161      & ~925       & 1093      & 1551      &&  $-$1712     &  1370     &  783      &&   1733     &  2032     \\
& [$7f$]        && 2156       & ~931       & 1098      & 1554      &&  $-$1739     &  1354     &  773      &&   1731     &  2037     \\
& [$8f$]        &&  2157      & ~929       & 1096      & 1550      &&  $-$1738     &  1350     &  770      &&   1730     &  2031     \\
& [$9f$]        &&  2163      & ~925       & 1091      & 1546      &&  $-$1759     &  1343     &  763      &&   1732     &  2027     \\
& [$10f$]       &&  2152      & ~926       & 1094      & 1549      &&  $-$1743     &  1349     &  768      &&   1727     &  2030     \\
 &          &&            &           &           &           &&       &           &           &&            &           \\
MR-CI &   [$10f$]   &&  2114      & ~930       & 1103      & 1561      &&  $-$1682     &  1374     &  788      &&   1711     &  2041     \\
MR-CI-red & [$10f$]  &&  2119      & ~929       & 1102      & 1559      &&            &           &           &&            &           \\
 &          &&            &           &           &           &&       &           &           &&            &           \\
MR-MCHF & $\{ 10s9p8d4f \}$  && 2122      & ~930       & 1102      & 1560      &&  $-$1697     &  1369     &  784      &&   1715     &  2040     \\ [0.1cm]
\hline  \\[-0.2cm]
BP & [$8f$]     &&  2102      & ~839       & 1099      & 1543      &&  $-$1616     &  1742   & 1007      &&   1763     &  1685     \\ 
& [$9f$]     &&  2110      & ~835       & 1093      & 1538      &&  $-$1637     &  1737     & 1001      &&   1772     &  1686     \\ 
& [$10f$]     &&  2100      & ~837       & 1095      & 1542      &&  $-$1620     &  1745     & 1007      &&   1768     &  1691     \\
&          &&            &           &           &           &&            &           &           &&            &           \\
RCI-P &  [$8f$]   &&  2103      & ~839       & 1097      & 1540      &&  $-$1615     &  1737     & 1005      &&   1763     &  1686    \\
& [$9f$]    &&  2112      & ~835       & 1091      & 1536      &&  $-$1636     &  1732     &  999      &&   1772     &  1686    \\
& [$10f$]   &&  2101      & ~836       & 1093      & 1540      &&  $-$1620     &  1740     & 1005      &&   1768     &  1691    \\
&          &&            &           &           &           &&            &           &           &&            &           \\
MR-BP  & [$10f$] &&  2073      & ~846       & 1107      & 1553      &&  $-$1572          &    1762       &  1022        & &    1759       &  1701        \\
 &          &&            &           &           &           &&       &           &           &&            &           \\
          &           && \multicolumn{1}{c}{1857.1(2.1)} & \multicolumn{1}{c}{1746.5(1.5)} \\[0.1cm]
\hline \hline \\
\end{tabular}
\end{center}
\label{tab:A_J_MCHF_BP}
\end{table}

\begin{table}
\caption{Hyperfine structure constants $A_J$ (in MHz) of $2p^4(^3P)3p \; ^4D^o_{J}$, $\; ^4P^o_{J}$ and $\; ^2D^o_{J}$ obtained with the fully relativistic multiconfiguration-Dirac-Hartree-Fock method using single- and double-monoreference expansions (SD-MCDHF), and multireference relativistic configuration interaction (MR-RCI) calculations.}
\begin{center} 
\begin{tabular}{lcccccccccccc} \\
 \\
\hline \hline \\[-0.2cm]
    &  && \multicolumn{4}{c}{$^4D^o$} && \multicolumn{3}{c}{$^4P^o$}& \multicolumn{2}{c}{$^2D^o$ }  \\
Method & [AS]       &&$A_{1/2}$&$A_{3/2}$&$A_{5/2}$&$A_{7/2}$&&$A_{1/2}$&$A_{3/2}$& $A_{5/2}$ & $A_{3/2}$& $A_{5/2}$ \\ [0.1cm]
\hline  \\[-0.2cm]
$\{  2p^43p + 2p^5 \}$ &       && 2250    & ~809     & 1125    & 1529    &&   $-$1772  &  1811   & 1045      &  1795      &   1633      \\ 
&&&&&&&&&&&& \\ 
SD-MCDHF & [3]        & & 1380    & ~1086    & 1532    & 1883    &&   ~$-$317   &  2418   & 1525      &  1362      &   1914     \\
& [4]         && 2120    & ~746     & 1035    & 1461    &&   $-$1769  &  1713   &  ~948      &  1783     &   1630     \\
& [5]         && 2027    & ~785     & 1109    & 1533    &&   $-$1708  &  1790   &  ~995      &  1748     &   1673     \\
& [$6f$]        && 2087    & ~782     & 1083    & 1511    &&   $-$1663  &  1777   & ~992      &  1788     &   1650     \\
& [$7f$]      && 2061    & ~805     & 1131    & 1526    &&   $-$1601  &  1818   & 1023      &  1818     &   1739    \\
& $ \{8s7p6d4f\}$        && 2037    & ~814     & 1127    & 1529    &&   $-$1583  &  1807   & 1017      &  1814     &   1738     \\
& $ \{9s8p7d4f\}$     && 2065    & ~812     & 1109    & 1523    &&   $-$1624  &  1780   & ~998      &  1784     &   1660     \\
& $ \{10s9p8d4f\}$       && 2060    & ~818     & 1111    & 1526    &&   $-$1606  &  1784   & 1002      &  1789     &   1666     \\
&          & &        &         &         &         &       &   &         &           &           &             \\
MR-RCI &  $\{10s9p8d3f\}/[5d]$  && 2071    & ~850     & 1127    & 1546    &&   $-$1555  &  1784    & 1028      &  1772     &   1700     \\
       &   (see text)           &&         &          &         &         &&            &         &           &           &            \\
&          & &        &         &         &         &   &       &         &           &           &             \\
\hline \hline \\
\end{tabular}
\end{center}
\label{tab:A_J_MCDHF}
\end{table}

\begin{table}
\caption{Comparison of the hyperfine constants $A_J$ (in MHz) estimated from non relativistic (SD-MCHF, MR-MCHF and MR-CI) and relativistic calculations. From the non-relativistic MCHF orbitals, relativity is included in the Breit-Pauli approximation, mono-reference (BP) and multi-reference (MR-BP), or through relativistic configuration interaction calculations using one-electron orbitals built in the Pauli approximation (RCI-P). Fully relativistic multiconfiguration Dirac-Hartree-Fock (MCDHF)  and multireference relativistic configuration interaction (MR-RCI)  are also reported and compared with observation.}
\begin{center} 
\begin{tabular}{llrrrrrrrrrrrcrrrrrrrrc} \\
 \\
\hline \hline \\[-0.2cm]
& & &  \multicolumn{3}{c}{Non relativistic} & & \multicolumn{5}{c}{Relativistic} & \\
& & &  \multicolumn{1}{c}{mono-} & \multicolumn{2}{c}{multi-reference} & & \multicolumn{3}{c}{mono-} & \multicolumn{2}{c}{multi-reference}  &
& Observed  \\
& & & & & & & & & & & \\
   \multicolumn{1}{c}{Term}            & $A_J$   & \hspace*{0.5cm} & SD-MCHF  & MR-MCHF & MR-CI &\hspace*{0.5cm} & BP &   RCI-P  & MCDHF & MR-BP  & MR-RCI & \hspace*{0.5cm} &  \\  [0.1cm]
\hline  \\[-0.2cm]
$2p^4(^3P)3p~^{4}D^o$ & $A_{1/2}$ && 2152~~~    & 2122~~~    & 2114  && 2100 & 2101 & 2060 & 2073    &  2071   &&           \\
                                         & $A_{3/2}$ &&  926~~~    &  930~~~    &  930  && 837 &   836 &  818 & 846    & 850   &&           \\
                                         & $A_{5/2}$ && 1094~~~    & 1102~~~    & 1103  && 1095 &  1093 & 1111 & 1107    & 1127   && 1148(1)~\cite{Levetal:2007a}   \\
                                         & $A_{7/2}$ && 1549~~~    & 1560~~~    & 1561  && 1542 &  1540 & 1526 & 1553    &1546   && 1564(1)~\cite{Levetal:2007a}   \\
&&&&&&&&&&&&&& \\
$2p^4(^3P)3p~^{4}P^o$ & $A_{1/2}$ && $ -$1743~~~    &$-$1697~~~    &$-$1682  && $-$1620 & $-$1620 &$-$1606 & $-$1572    & $-$1555   &&           \\
                                         & $A_{3/2}$ && 1349~~~    & 1369~~~    & 1374  && 1745 &   1735 & 1784  & 1762    & 1784  &&           \\
                                         & $A_{5/2}$ &&  768~~~    &  784~~~    &  788  &&  1007 &  1004 & 1002 & 1022    & 1028  &&    \\
&&&&&&&&&&&&&& \\   
 $2p^4(^3P)3p~^{2}D^o$ & $A_{3/2}$ && 1727~~~    & 1715~~~    & 1711  && 1768 &  1768 & 1789 & 1759    & 1772   && 1857.1(2.1)~\cite{TatAtu:97a} \\
                                         & $A_{5/2}$ && 2030~~~    & 2040~~~    & 2040  &&  1691 &  1691 & 1666 & 1701    & 1700   && 1746.5(1.5)~\cite{TatAtu:97a} 
                                          \\[0.1cm]
\hline \hline \\\end{tabular}
\end{center}
\label{tab:A_J_comparison}
\end{table}

\begin{table}
\caption{Comparison of the orbital, spin-dipole and contact contributions to the hyperfine structure constants (all numbers in MHz) calculated with the non relativistic SD-MCHF method and including the relativistic Breit-Pauli corrections (BP). The $\Delta A_{i}/A_{tot}$ contributions are defined in the text (see eq.~(\ref{A_i_over_A_tot})).}
\begin{center}
\begin{tabular}{llrrrrrrrrrrrrrrrrrrcccccccccccc} \\
 \\
\hline \hline \\[-0.2cm]
 &         & \multicolumn{3}{c}{$J=1/2$ }      & \multicolumn{3}{c}{$J=3/2$ }     &    \multicolumn{3}{c}{$J=5/2$}  &\multicolumn{3}{c}{$J=7/2$}\\  [0.1cm]
\hline  \\[-0.2cm]
 &         & SD-MCHF & BP  &$\Delta A_{i}/A_{tot}$& SD-MCHF & BP &$\Delta A_{i}/A_{tot}$& SD-MCHF& BP &$\Delta A_{i}/A_{tot}$& SD-MCHF & BP &$\Delta A/A_{tot}$\\  [0.1cm]
\hline  \\[-0.2cm]
$^{4}D^o$  &A$_{orb}$& 4360  &4028 &  $-$15.4~\%             & 1744  &1562& $-$19.7~\%              & 1370 &1287&   $-$7.6~\%             &  1246 &1246&     0~\%          \\
 &A$_{sd}$ &$-$2099  &$-$1821&   12.9~\%             & $-$840  &$-$759&   8.7~\%              & $-$317 &$-$244&    6.7~\%             &   257 & 248&  $-$0.6~\%          \\
 &A$_{c}$  & $-$109  &$-$107 &    0.1~\%             &   22  &  34&   1.3~\%              &   40 &  52&      1~\%             &    47 &  47&     0~\%          \\
 &&&&&&&&&&& \\
 &A$_{tot}$& 2152  &2100 &   $-$2.4~\%             &  926  & 837&  $-$9.6~\%              & 1094 &1095&      0~\%             &  1549 &1542&  $-$0.5~\%          \\
 &         &       &     &                      &       &    &                      &      &    &                      &       &    &                  \\
    &&&&&&&&&&& \\ 
$^{4}P^o$&A$_{orb}$&$-$1454  &$-$1159&   16.9~\%             &  582  &1012&  31.9~\%              &  872 & 983&   14.5~\%             &        &    &                  \\
 &A$_{sd}$ & $-$497  & $-$668&   $-$9.8~\%             &  676  & 641&  $-$2.6~\%              & $-$179 & $-$51&   16.7~\%             &       &    &                  \\
 &A$_{c}$  &  208  &  207&      0~\%             &  91.5 &  92&     0~\%              &   75 &  75&      0~\%             &       &    &   
  &&&&&&&&&&& \\               \\
 &A$_{tot}$&$-$1743  &$-$1620&      7~\%             &  1349 &1745&  29.3~\%              &  768 &1007&   31.1~\%             &       &    &                 \\ &         &       &     &                      &       &    &                      &      &    &                      &       &    &                  \\ 
    &&&&&&&&&&& \\ 
$^{2}D^o$ &A$_{orb}$& & & & 2618  & 2657&    2.2~\%             &  1745 &1691&  $-$2.7~\%              &      &    &                                     \\
 &A$_{sd}$ & & & & $-$848  & $-$836&    0.7~\%             &   242 & $-$32& $-$13.5~\%              &      &    &                                   \\
 &A$_{c}$  & & & & $-$43  &  $-$53&   $-$0.6~\%             &    43 &  32&  $-$0.5~\%              &      &    &                    \\
   &&&&&&&&&&& \\ 
 &A$_{tot}$& & & & 1727  &1768 &    2.4~\%             &  2030 &1691& $-$16.7~\%              &      &    &                  \\
    &&&&&&&&&&& \\ [0.1cm]
\hline \hline \\
\end{tabular}
\end{center}
\label{tab:A_J_ventilation}
\end{table}

\begin{table}
\caption{$J$-dependent  factors of the orbital (${\cal F}^{orb}(L,S,J)$), spin-dipole (${\cal F}^{sd}(L,S,J)$) and contact (${\cal F}^{c}(L,S,J)$)~contributions to the hyperfine constant $A_J$ (see equations~(\ref{sec3:F_orb}), (\ref{sec3:F_sd}) and (\ref{sec3:F_c})). $LS$ eigenvector compositions are given in \%.} 
\begin{center}
\begin{tabular}{l c r r r r r r r} \\
 \\
\hline \hline \\[-0.2cm]
$LS$ term & $J$     &   \multicolumn{1}{c}{\hspace*{0.5cm} ${\cal F}^{orb}$  }   &  \multicolumn{1}{c}{${\cal F}^{sd}$} 
 &   \multicolumn{1}{c}{${\cal F}^{c}$}  & \hspace*{1cm} & \multicolumn{3}{c}{composition (in \%)}  \\
\hline \hline \\[-0.2cm]
$2p^{4}3p~\; ^{4}D^o$  & $1/2$   &  $+35/35$   &   $- 245/105 $   & $  - 70/105$ & & $99.6 (^4D),$ & $0.3 (^4P),$ & $0.1 (^2P)$\\
&&&&&&&\\
                  & $3/2$   &   $+14/35$   & $ - 98/105 $   & $  + 14/105 $  && $96.9 (^4D),$ & $2.2 (^2D),$ & $0.8 (^4P)$ \\
&&&&&&&\\
                  & $5/2$   &  $ +11/35 $  & $ - 37/105$  & $   +26 / 105 $  && $92.5 (^4D),$ & $6.5 (^2D),$ & $0.9 (^4P)$ \\
&&&&&&&\\
                  & $7/2$   &  $+10/35$   &   $+ 30/105$   &    $+30/105$ && $100.0 (^4D)$ & & \\[0.1cm]
\hline \\[0.01cm]
$2p^{4}3p~\; ^{4}P^o$ & $1/2$  &   $- 10/15$   &   $+ 50/45$& $ + 50/45 $ && $99.6 (^4P),$ & $0.2 (^4D),$ & $0.2 (^2S)$ \\
&&&&&&&\\
                 & $3/2$   &  $ + 4/15 $&  $- 68/45$  &  $ + 22/45$ && $98.8 (^4P),$ & $0.7 (^4D),$ & $0.5 (^4S)$\\
&&&&&&&\\
                 & $5/2$   &   $+ 6/15$   & $ + 18/45$  & $ + 18/45 $ && $98.9 (^4P),$ & $1.0 (^4D),$ & $0.1 (^2D)$\\[0.1cm]
\hline \\[0.01cm]
$2p^{4}3p~\; ^{2}D^o$  & $3/2$   &   $+ 3/5$   & $ - 7/5$   & $  - 2/5 $ && $96.9 (^2D),$ & $2.7 (^4D),$ & $0.3 (^2P)$ \\
&&&&&&\\
                  & $5/2$   &  $ + 2/5 $  & $ + 2/5 $  & $   + 2/5 $ && $91.4 (^2D),$ & $8.1 (^4D),$ & $0.5 (^4P)$\\[0.1cm]
\hline
\hline
\end{tabular}
\end{center}
\label{tab:F_factors}
\end{table}

\clearpage

\section*{ACKNOWLEDGMENTS}

This work was supported by the Direction G\'en\'erale de la Recherche Scientifique et du D\'eveloppement Technologique (DGRSDT) of Algeria, the Communaut\'e fran\c{c}aise  of Belgium (Action de Recherche Concert\'ee), the Belgian National Fund for Scientific Research (FRFC/IISN Convention) and by the IUAP Belgian State Science Policy (Brix network P7/12).

\bibliography{F_I_hfs.bib}

\end{document}